\let\accentvec\vec 
\documentclass{aa}

\let\vec\accentvec
\usepackage{amsmath}
\usepackage{amsmath}
\usepackage{txfonts}
\usepackage{balance}
\usepackage{natbib}
\bibpunct{(}{)}{;}{a}{}{,} 
\usepackage{xspace}
\usepackage[normalem]{ulem} 
\usepackage{multirow}
\usepackage{rotating}
\usepackage{bbm}
\usepackage{cancel}
\usepackage{graphicx}
\usepackage{mathabx}
\usepackage{ifthen}
\usepackage{booktabs}
\usepackage{tabularx}
\usepackage{dcolumn}
\usepackage{float}

\newcommand{\del}{\partial}

\newcommand{\St}{\text{St}\xspace}
\newcommand{\rhos}{\ensuremath{\rho_\text{s}}\xspace}
\newcommand{\rhodust}{\ensuremath{\rho_\mathrm{d}}\xspace}
\newcommand{\rhogas}{\ensuremath{\rho_\mathrm{g}}\xspace}

\newcommand{\uf}{\ensuremath{u_\text{f}}\xspace}
\newcommand{\csound}{\ensuremath{c_\mathrm{s}}\xspace}

\newcommand{\Ok}{\ensuremath{\Omega_\mathrm{k}}\xspace}

\newcommand{\Siggas}{\ensuremath{\Sigma_\mathrm{g}}\xspace}
\newcommand{\Sigdust}{\ensuremath{\Sigma_\mathrm{d}}\xspace}

\newcommand{\alphat}{\ensuremath{\alpha_\text{t}}\xspace}
\newcommand{\alphaD}{\ensuremath{\alpha_\text{D}}\xspace}
\newcommand{\alphaA}{\ensuremath{\alpha_\text{A}}\xspace}

\newcommand{\Rc}{\ensuremath{r_\mathrm{c}}\xspace}
\newcommand{\Rd}{\ensuremath{r_\mathrm{dz}}\xspace}
\newcommand{\Edr}{\ensuremath{E_\mathrm{d}}\xspace}
\newcommand{\kap}{\ensuremath{\kappa_{880}}\xspace}
\newcommand{\dlnPdlnR}{\ensuremath{ {\left| \frac{\mathrm{d}\ln P}{\mathrm{d}\ln r} \right| } }\xspace}
\usepackage[
pdfstartview=FitH, linkcolor=blue, anchorcolor=blue, citecolor=blue, filecolor=blue, menucolor=blue,
urlcolor=blue, colorlinks=true,breaklinks]{hyperref}

\usepackage{pdfsync}

\makeatletter \if@referee
    \usepackage{endfloat}
    
\else
    
\fi \makeatother

\defcitealias{Birnstiel:2012p17135}{BKE12}

\begin{document}
\title{Can grain growth explain transition disks?}
\titlerunning{Can grain growth explain transition disks?}
\author{T. Birnstiel\inst{1}\fnmsep\inst{2} \and S. M. Andrews\inst{3} \and B. Ercolano\inst{1}\fnmsep\inst{2}}
\authorrunning{T.~Birnstiel et al.}
\institute{
    University Observatory Munich, Scheinerstr. 1, D-81679 M\"unchen, Germany \and Excellence
    Cluster Universe, Technische Universit\"at M\"unchen, Boltzmannstr. 2, 85748 Garching, Germany
    \and Harvard-Smithsonian Center for Astrophysics, 60 Garden Street, Cambridge, MA 02138, USA
}
\date{\today}

\abstract
{}
{Grain growth has been suggested as one possible explanation for the diminished dust
optical depths in the inner regions of protoplanetary ``transition" disks.  In this work, we
directly test this hypothesis in the context of current models of grain growth and transport.}
{A set of dust evolution models with different disk shapes, masses, turbulence parameters, and drift
efficiencies is combined with radiative transfer calculations in order to derive theoretical
spectral energy distributions (SEDs) and images.}
{We find that grain growth and transport effects can indeed produce dips in the infrared SED, as
typically found in observations of transition disks. Our models achieve the necessary
reduction of mass in small dust by producing larger grains, yet not large enough to be fragmenting
efficiently.  However, this population of large grains is still detectable at millimeter
wavelengths. Even if perfect sticking is assumed and radial drift is neglected, a large
population of dust grains is left behind because the time scales on which they are swept up by the
larger grains are too long. This mechanism thus fails to reproduce the large emission
cavities observed in recent millimeter-wave interferometric images of accreting transition disks.}
{}


\maketitle

\section{Introduction}\label{sec:introduction}
The evolution of circumstellar disks, the birthplaces of planets, is still enigmatic, even though
the pioneering theoretical work on this topic was started almost 40 years ago by
\citet{LyndenBell:1974p1945}.  Viscous and/or gravitational stresses are the drivers of the disk
accretion flows, which can be traced indirectly by the observed disk lifetimes and accretion rates
\citep[e.g.,][]{Hartmann:1998p664,SiciliaAguilar:2006p10361,Hernandez:2007p4281,Fedele:2010p15973}.

While the general trends of declining accretion rates and disk masses can be explained by viscous
evolution, a sub-set of objects called transition disks remain mysterious. These objects appear
dust-depleted in their inner regions, while the outer regions resemble normal circumstellar disks
\citep{Strom:1989p9475, Skrutskie:1990p16132, Calvet:2002p10424, Espaillat:2007p17013,
Espaillat:2010p17008, Andrews:2011p16142}.
 The sizes of the dust cavities range from a few to more than 70~AU
\citep[e.g.,][]{Pietu:2006p17014, Hughes:2007p17475, Brown:2008p8893, Hughes:2009p17047,
Andrews:2009p7729, Brown:2009p8895, Isella:2010p9438, Andrews:2010p17519, Isella:2010p17527,
Andrews:2011p16142}.
 The gas content of the cavities is still largely unknown.  It may also be reduced compared to the
outer disk regions as found for example by \citet{Najita:2010p17082}, \citet{Dutrey:2008p17530}, or
\citet{Lyo:2011p16794}; yet other works such as \citet{Pontoppidan:2008p9993} or
\citet{Salyk:2011p17115} do detect gas inside the dust cavities.

The time scale of this transition phase of disk evolution is estimated to be of the order of a few
times $10^5$ years \citep{Skrutskie:1990p16132,Hartigan:1990p16133}.  This estimate, however, is
based on the transition disk frequency, which is still only a lower limit due to the lack of spatial
resolution. Classifying transition disks only based on the SED can be misleading because steep
decreases in the dust surface density can easily be missed due to the presence of small dust
particles. Some mechanisms proposed to explain observations of transition disks include planet-disk
interactions \citep[e.g.,][]{Rice:2003p15994,Zhu:2011p16181}, or photo-evaporation
\citep{Clarke:2001p969,Alexander:2006p136,Ercolano:2008p13616}. Recent imaging of disks by
\citet{Andrews:2011p16142} revealed a higher fraction ($>20$\%) of disks with large cavities for the
mm-brightest sources. Higher fractions of transition disks, indicating longer disk clearing time
scales, are more difficult to explain by photoevaporation. However, the dust emission signature of
photoevaporating disks has not yet been self-consistently modeled, treating grain growth physics and
dust-gas feedback. It is therefore unclear whether photoevaporating disks leave a dust rich or a
completely dust and gas free cavity behind \citep[e.g.,][]{Alexander:2007p131,Garaud:2007p405}.

The formation of large disk cavities by viscous evolution is generally problematic, even
with the assumption that photoevaporation or some other mechanism is able to decouple the outer and
inner disc at larger radii. Indeed the viscous time at 35~AU for typical values for the viscosity
parameter\footnote{Assuming $\alphat = 3 \times 10^{-3}$. Both observational and theoretical works
indicate values between $10^{-3}$ and $10^{-2}$.} is 3.4~Myr.
On the other hand, even at 35~AU, the time scales for grain growth and radial drift are only a
couple of thousand years. Any mechanism that triggers significant changes in the dust evolution
(e.g., the emergence of a pressure maximum, see \citealp{Pinilla:2012p16999}) could therefore
quickly induce observational signatures. The fact that grains grow and consequently become more
mobile due to radial drift is likely an important part of the solution to this problem.

Recent observations by \citet{Kraus:2012p16079} found a possible planetary-mass companion inside the
cavity of the LkCa 15 transition disk. It remains to be shown whether planets cause gaps/pressure bumps
or vice versa 
(e.g., \citealp{Kretke:2007p697,Brauer:2008p212}; Pinilla et al., in prep.).
Determining the origin of transition disks is therefore one of the most fundamental issues in our 
efforts to forge a better understanding of planet formation.

It has been suggested, although not demonstrated in any detail, that the growth and radial transport
of dust could potentially explain the observed signatures of transition disks with large inner holes
\citep[e.g.,][]{Dullemond:2005p378,Tanaka:2005p6703,Najita:2008p17130,Pontoppidan:2008p9993}. The
observations indicate a decreased {\it optical depth} in the inner regions of transition disks
both in the IR as well as in mm-observations, which does not necessarily imply
diminished dust densities. Grain growth therefore offers two (related) pathways towards this end.
First, particles decouple from the gas as they grow, which causes them to spiral
inwards \citep{Weidenschilling:1977p865,Nakagawa:1986p2048}. This way, the dust optical depths are reduced
by the actual removal of dust mass. Second, grain growth itself causes a decrease in
the dust opacities, since larger particles emit less efficiently, which also decreses the
otpical depth.
Dust grains of sizes beyond a few centimeters become basically invisible to observations. However, the edges of transition disks seem
to be relatively abrupt; they have been modeled as step functions
\citep{Andrews:2011p16142,Isella:2012p17671} or steep power-law profiles
\cite{Isella:2010p17527,Isella:2012p17671}. This suggests that the environment for grain evolution
must abruptly change at this point in the disk. Pressure bumps or even gaps opened by planets
might be possible explanations \citep{Lin:1986p12877,Zhu:2011p16181}. Another possibility could
be the outer edges of dead-zones \citep{Gammie:1996p1515}. Dead zones are regions where the
ionization fraction of the disk drops below the critical value needed to drive the
magneto-rotational instability \citep[MRI,][]{Balbus:1991p4932}, the widely accepted source of
turbulent viscosity.

In this work, we want to investigate whether grain growth and transport -- either alone or aided by
a dead zone -- can be the cause of the observed transition disk signatures. In
Section~\ref{sec:background} we will discuss some of the equations which are crucial for the
understanding of our modeling results. Section~\ref{sec:model} describes the model setups and
assumptions. The resulting simulation outcomes and simulated observations are shown and explained in
Section~\ref{sec:results}. Our findings are discussed and summarized in Section~\ref{sec:summary}.

\section{Background}\label{sec:background}
In the following section, we will summarize some of the results of \citet[][hereafter
\citetalias{Birnstiel:2012p17135}]{Birnstiel:2012p17135}, which motivated the simulations presented
in this paper and will help to understand the results.
It was found in \citetalias{Birnstiel:2012p17135} that the upper end of the grain size distribution
can be limited by two effects, namely fragmentation and radial drift. In the
former case, due to the fact that typical impact velocities increase with grain size, the grains can
only grow until they reach a size at which fragmentation sets in. The size limit for turbulent
relative velocities was found to be
\begin{equation}
a_\mathrm{frag} \approx 0.08 \, \frac{\Siggas}{\rhos \alphat} \, \frac{\uf^2}{\csound^2}.
\label{eq:a_frag}
\end{equation}
Here, \uf denotes the fragmentation threshold velocity and \rhos the specific density of the dust
grains. \Siggas is the gas surface density, \csound the sound speed and \alphat the turbulent
viscosity parameter \citep{Shakura:1973p4854}. The other size limit is due to radial drift: grains
can only exist at a given radius if growth from smaller sizes resupplies them as fast as radial
drift removes them. In this case, the upper end of the size distribution can be approximated by
\citepalias[see][]{Birnstiel:2012p17135}
\begin{equation}
a_\mathrm{drift} \approx 0.35 \, \frac{\Sigdust}{\rhos\,\gamma}\left(\frac{r \, \Ok}{\csound}
\right)^2,
\label{eq:a_drift}
\end{equation}
where \Sigdust is the dust surface density, \Ok the Keplerian frequency and $\gamma$ the absolute
value of the power law index of the gas pressure $P$,
\begin{equation}
\gamma = \Edr\,\dlnPdlnR,
\label{eq:gamma}
\end{equation}
and we have introduced a parameter for the efficiency of radial drift, \Edr where $\Edr=1$
corresponds to the fiducial literature value. In the limit $a_\mathrm{frag}<a_\mathrm{drift}$,
fragmentation of the dust grains is the relevant size limit. In this case, small dust is constantly
resupplied in the form of fragments. In the opposite case, when the radial drift barrier is the
growth limiting factor, grains are removed by radial drift even before they reach sizes at which
fragmentation starts to become important.

From Eqns.~\ref{eq:a_frag} and \ref{eq:a_drift}, it can be seen that the quantities which most 
clearly determine the time evolution of the size limits are \Siggas and \Sigdust, respectively,
because the other quantities are not expected to change by orders of magnitude with time
at a given radius. However, it should be noted that in this study, we do not track the viscous
evolution of the gas surface density. This would lead to a further reduction of the maximum grain
size. A fixed gas surface density is therefore in tune with having a best-case scenario to test the
potential of the proposed mechanism. The fact that radial drift causes the dust surface density to
decrease faster than the gas surface density, even if viscous evolution were to be
included, means that the drift limit can become more important as the disk evolves.

The dependence of Eqns.~\ref{eq:a_frag} and \ref{eq:a_drift} on \Siggas and \Sigdust also causes
both size limits to decrease with radius, as observed by \citet{Banzatti:2011p15323} and
\citet{Guilloteau:2011p15287}. If the largest grain size at a given radius is limited by radial
drift, dust collision velocities are not high enough to cause fragmentation
\citepalias{Birnstiel:2012p17135}. The drifting grains therefore sweep up the small dust without
resupplying it by fragmentation. This means that drift-limited grain growth could naturally lead to
conditions where the inner regions are devoid of small dust (due to lack of fragmentation), while
small grains are still present in the outer disk (due to lower grain size limits in the outer
regions).

The most important and uncertain parameters are \uf, the collision velocity at which fragmentation
sets in, and the turbulent state of the disk, which is described by the parameter \alphat.
Estimates for \uf range from a few up to 35 m s$^{-1}$  \citep[see,][and references
therein]{Wada:2008p4903,Paszun:2009p8871,Blum:2008p1920}. We use values of 3 and 10~m~s$^{-1}$,
because icy aggregates are expected to fragment only at larger velocities compared to the
$\sim$1~m~s$^{-1}$, found for silicate dust grains \citep{Blum:1993p4324,Blum:2008p1920}.

\begin{figure}[tb]
  \centering
  \resizebox{\hsize}{!}{\includegraphics{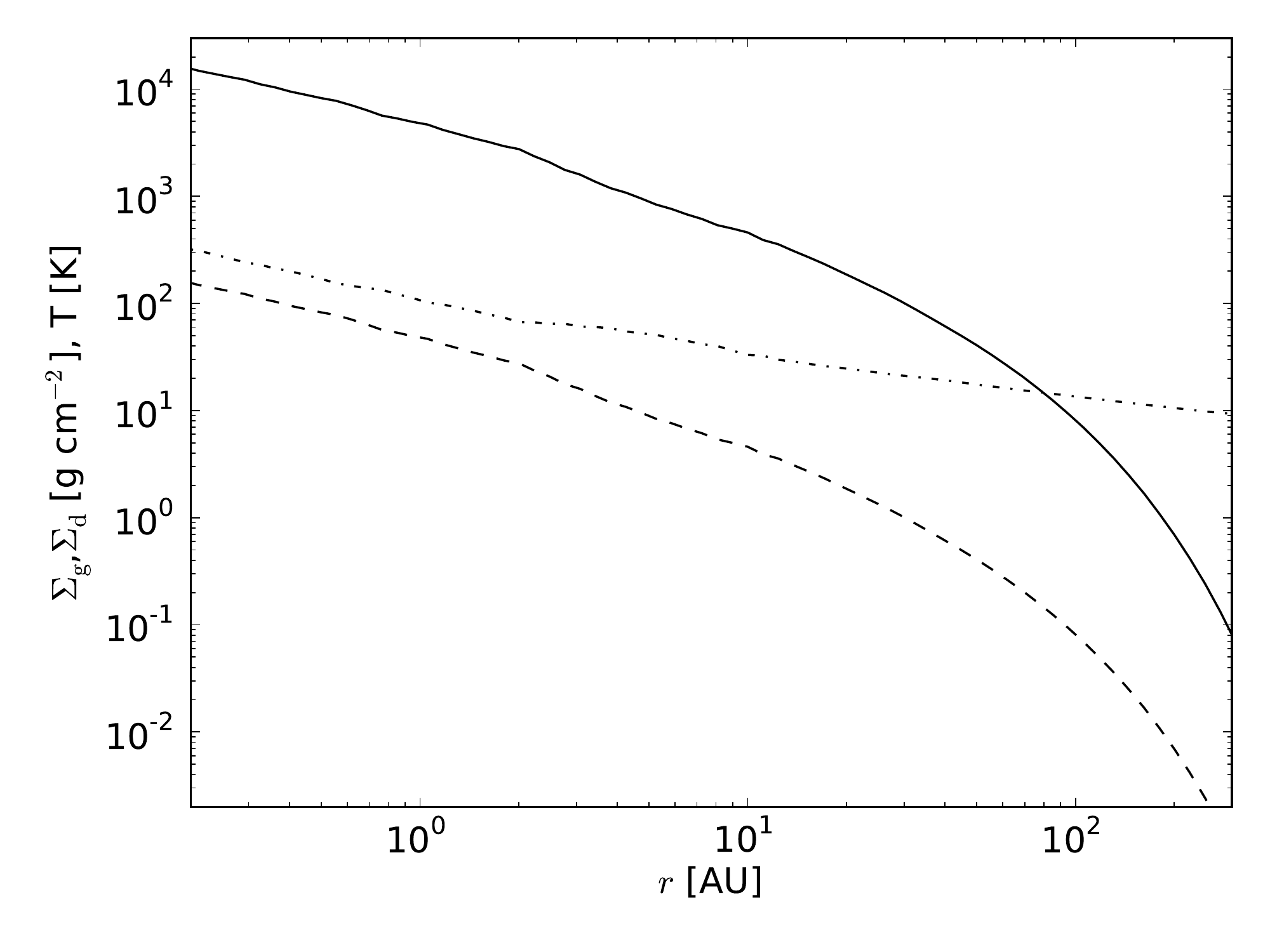}}
  \caption{Initial conditions of the simulations without a dead zone. Solid and dashed
  lines denote the gas and dust surface density in g cm$^{-2}$, respectively. The dash-dotted line
  denotes the mid-plane temperature.}
  \label{fig:initial_condition}
\end{figure}

\section{Modeling approach}\label{sec:model}

\begin{figure}[tb]
  \centering
    \makeatletter \if@referee
        \resizebox{0.6\hsize}{!}{\includegraphics{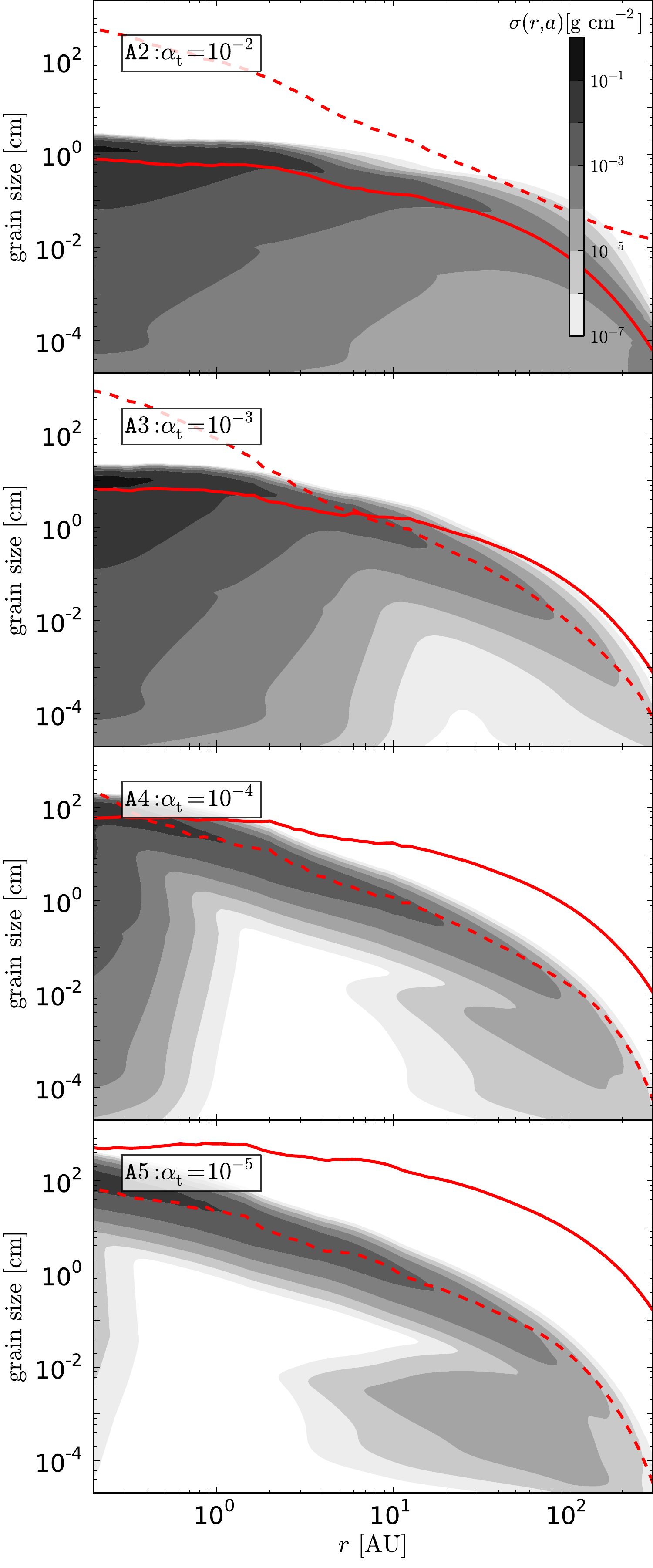}}
    \else
        \resizebox{0.9\hsize}{!}{\includegraphics{plots/contour_active_final.pdf}}
    \fi \makeatother
  \caption{Vertically integrated dust surface density distribution after 5 Myrs of evolution for
  the simulations without a dead zone, \texttt{A2} (top) to \texttt{A5} (bottom).
  The parameters for the simulations are shown in Table~\ref{tab:initial_conditions}.
  The solid red line denotes the growth barrier set by grain fragmentation, the dashed red line the size
  limit due to radial drift.}
  \label{fig:contours_active}
\end{figure}

\begin{figure}[tb]
  \centering
    \makeatletter \if@referee
        \resizebox{0.6\hsize}{!}{\includegraphics{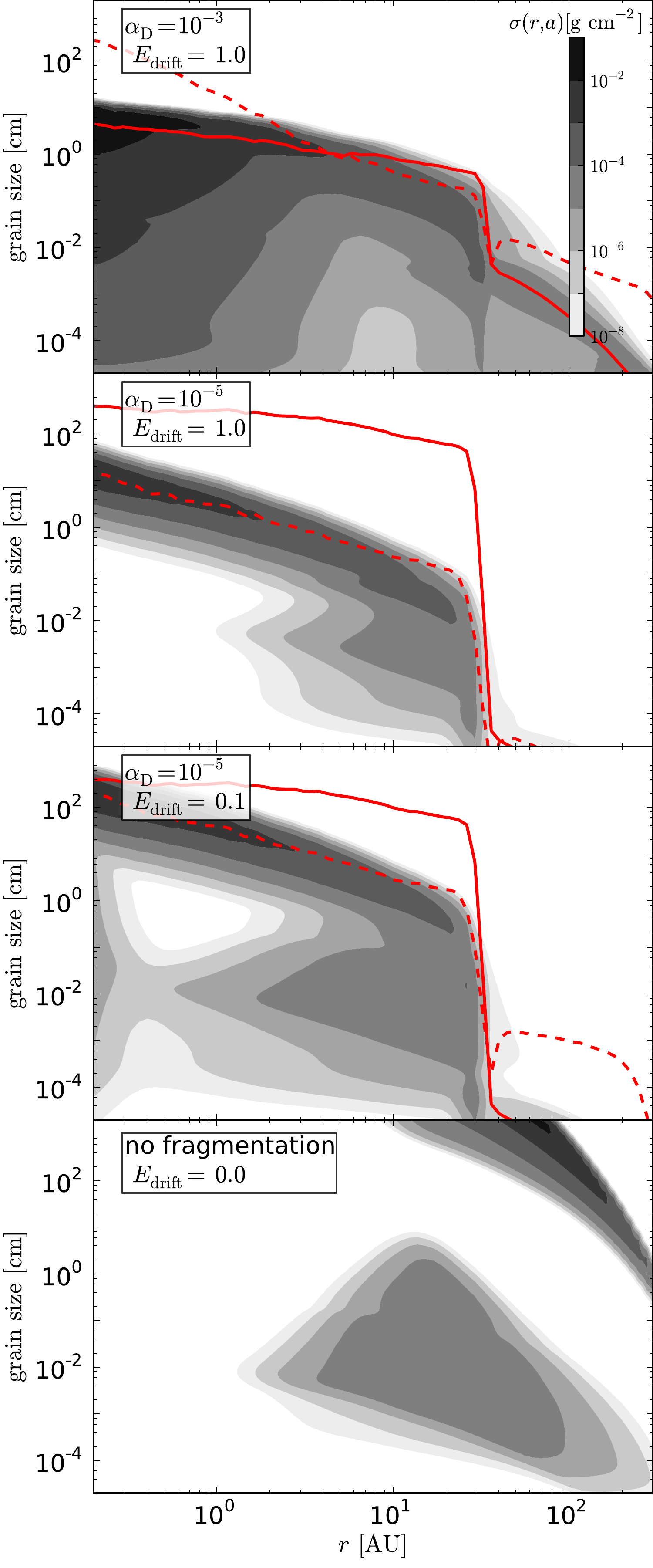}}
    \else
        \resizebox{0.9\hsize}{!}{\includegraphics{plots/contour_dead_other.pdf}}
    \fi \makeatother  
  \caption{Vertically integrated dust surface density distribution after 5 Myrs of evolution for
  simulations \texttt{D23\_M05} (top), \texttt{D25\_M05} (second panel), \texttt{LDE}
(third panel) and \texttt{COAG} (bottom panel). The corresponding parameters are shown in
Table~\ref{tab:initial_conditions}. The solid red line denotes the growth barrier set by grain
fragmentation, the dashed red line the size limit due to radial drift.}
  \label{fig:contours_dead}
\end{figure}

\subsection{Dust evolution}\label{sec:model:dustevol}
In this work, we use a vertically averaged dust evolution code which tracks the radial and size
evolution of the dust surface density of the disk. Effects of coagulation, fragmentation, and cratering
as well as radial drift, gas drag, and turbulent mixing are taken into account. For any details of
the dust evolution model, we refer to \citet{Birnstiel:2010p9709}.

We model both active disks, where the complete disk is MRI active as well as disks with MRI-inactive
regions, the so-called dead zones. The properties and even the mere existence of dead zones is still
an active matter of debate. In order to investigate whether a dead-zone can produce the
observational appearance of a protoplanetary disk, we have to make some simplifications and
assumptions. Therefore, we include the two most prominent effects of a dead zone which can alter the
evolution of dust, namely the decreased amount of turbulence and increased gas surface density
in the MRI inactive regions.

For the gas disk, we use a fixed surface density profile with a constant mass accretion rate up to a
characteristic radius $\Rc$,
\begin{equation}
\Siggas(r) \propto \frac{\Ok}{\alphat \, \csound^2} \cdot \left(1- \sqrt{\frac{r_\mathrm{in}}{r}}
\right) \cdot \exp\left(- \frac{r}{\Rc}\right),
\label{eq:gas_profile}
\end{equation}
where $r_\mathrm{in}$ is the innermost radius of the disk. In order
to mimic the effects of a dead zone, we use an alpha profile of
\begin{equation}
\alphat(r) = \alphaD - (\alphaD - \alphaA) \cdot \left\{
\begin{array}{ll}
\frac{1}{2} \exp\left( \frac{r-\Rd}{\Delta r} \right) & \text{if } r \leq \Rd\\
1 -         \frac{1}{2} \exp\left(-\frac{r-\Rd}{\Delta r} \right) & \text{if } r > \Rd
\end{array}\right.
\label{eq:alpha_profile}
\end{equation}
which represents a smooth transition at $\Rd$ from the turbulence parameter in the dead zone \alphaD
to the active one \alphaA over a transition width $\Delta r$, which we arbitrarily set to 1~AU. This
jump in \alphat causes also a jump in the gas surface density profile, as can be seen from
Eq.~\ref{eq:gas_profile}.

The initial condition for the dust surface density is given by a constant dust-to-gas ratio in
$\mu$m sized grains. The temperature profile is derived from two-dimensional radiative transfer
calculations (see Sect.~\ref{sec:model:RT}) and is kept fixed throughout the simulation. As an
example, the initial gas and dust surface densities and the initial temperature profile of
simulation \texttt{A2} is shown in Fig.~\ref{fig:initial_condition}.

\subsection{Vertical structure}\label{sec:model:vertical_structure}
Following the results from \citet{Dubrulle:1995p300}, the grain evolution code used in this work
assumes a Gaussian vertical distribution for the dust, where the scale heights depend on the grain size 
\citep[see][]{Birnstiel:2010p9709}. This assumption is very accurate
at the mid-plane, but it deviates in the surface layers of the disk. It can therefore be used for
the dust size evolution, which mostly depends on the mid-plane values. However, to derive proper
observables from the simulation outcome, the two-dimensional dust density distribution needs to be
reconstructed from the dust surface densities.

To this end, the vertical structure is assumed to be stationary and isothermal and we solve
numerically for the equilibrium between vertical mixing and dust settling,
\begin{equation}
\frac{\del \rhodust}{\del t} = - \frac{\del}{\del z} \left[ \rhodust \, u_\mathrm{sett} - 
D_\mathrm{d} \, \rhogas \frac{\del}{\del z} \left(\frac{\rhodust}{\rhogas}\right) \right] = 0,
\label{eq:settling_mixing}
\end{equation}
where \rhodust and \rhogas are the dust and gas densities,
\begin{equation}
D_\mathrm{d} = \frac{\nu_\mathrm{g}}{1+\St^2}
\label{eq:dust_diffusivity}
\end{equation}
is the dust diffusivity \citep{Youdin:2007p2021}; we assume that the gas diffusivity equals the
gas viscosity $\nu_\mathrm{g}$. \St is the particle Stokes number and
\begin{equation}
u_\mathrm{sett} = \frac{3\sqrt{\pi}\, m \, z \, \Ok^2}{8\sqrt{2}\rhogas \, \pi \, a^2 \, \csound}
\label{eq:u_sett}
\end{equation}
is the settling velocity for spherical grains in the Epstein regime \citep{Nakagawa:1986p2048}. We
denote the grain masses and radii as $m$ and $a$, respectively. The resulting two-dimensional
density distribution of each grain size is then used in the radiative transfer calculations.

\subsection{Radiative transfer calculations}\label{sec:model:RT}
The dust evolution models described above need to be converted into synthetic data products to
assess how well they can reproduce the key observational properties of transition disks.  For each
model, we compute a broadband SED and high resolution millimeter-wave continuum image following the
basic procedures outlined by \citep{Andrews:2009p7729,Andrews:2011p16142}.  The dust density
structure $\rho_d$ for each grain size $a$ was re-sampled onto a fixed grid in spherical
coordinates, with high resolution refinements near the disk midplane and around regions of strong
optical depth gradients (i.e., at the inner edge near 0.1 AU and the outer boundary of the dead
zone).  Absorption and scattering opacities for each grain size were calculated with a Mie code,
assuming a population of segregated spheres with the optical constants and material compositions
advocated by \citet{Pollack:1994p9428} \citep[see also][]{Andrews:2012p16676}.  We assume that
stellar irradiation is the only relevant heating source, and set the stellar photosphere to have
properties that are representative of typical transition disk hosts: $T_{\rm eff} = 4300$ K,
$R_{\ast} = 2.5$ R$_{\odot}$ (implying $L_{\ast} \approx 2$ L$_{\odot}$), and $M_{\ast} = 1$
M$_{\odot}$.

The two-dimensional Monte Carlo radiative transfer code {\tt RADMC} \citep{Dullemond:2004p380} was
used to simulate the propagation of radiation through each model structure and compute an
internally-consistent temperature structure.  That process was iterated with the dust evolution code
for each model so adjustments could be made to the vertical distribution of particles of a given
size.  We adopted a simple convergence criterion such that the midplane temperatures did not change
more than $\sim$10\%\ between iterations (in practice, this amounted to only 1 or 2 iterations).  A
ray tracing algorithm was then used on the final model structure to construct a theoretical SED and
880\,$\mu$m continuum image for a given disk viewing geometry and distance.
For simplicity, we fixed representative values for the latter; $d = 140$\,pc, a disk inclination $i
= 35\degr$, and major axis position angle of 155\degr\ (measured east of north).
The 880~$\mu$m model image is used to synthesize a simulated interferometric image like those
observed by the Submillimeter Array \citep[for examples, see][]{Andrews:2011p16142}.

\subsection{Shortcomings of the modeling approach}\label{sec:model:shortcomings}
A vertically integrated coagulation code combined with a vertical settling-mixing equilibrium
calculation is a good approximation of the vertical dust structure, as long as the disk is not
layered, as it is in the presence of a dead zone. For example, a small amount of the mid-plane dust
could reach the surface layers and be fragmented due to the high degree of turbulence in the active
regions. Furthermore, turbulent transport in these regions can be much stronger than in the
mid-plane \citep{Turner:2010p15252}. Due to the fact that our model only uses the (possibly very
low) mid-plane turbulence parameter, this model represents a 'best case' scenario in favor of the
removal of small grains.
For simplicity, the evolution of the gas surface density and also the dependence of the MRI
turbulence on the dust distribution \citep[e.g.,][]{Sano:2000p9889} are not treated in this work.

It is important to note that other obstacles for particle growth at smaller sizes have been
suggested, e.g., bouncing \citep{Guttler:2010p9745,Zsom:2010p9746} or charging effects
\citep{Okuzumi:2009p7473}. Observations, however, do find ample evidence of the presence of both
large \citep{Testi:2001p9427,Natta:2004p3169,Rodmann:2006p8905,Ricci:2010p9423} and small grains
\citep[e.g.,][and references therein]{Watson:2007p17366} in protoplanetary disks, which suggest that
these barriers have been overcome. The question remains whether the radial drift and the
fragmentation barrier are indeed the upper limits of the size distribution. At least fragmentation
seems necessary in order to explain the observed presence of small dust
\citep[see][]{Dullemond:2005p378}, however the expected rate of radial drift is too high to account
for the observed disk lifetimes \citep{Weidenschilling:1977p865,Brauer:2007p232}. On the other hand,
different sizes as well as different profiles of the dust and the gas disk
\citep{Panic:2009p11789,Andrews:2012p16676} are indicating that to some extent radial drift is at
work in circumstellar disks. Possibly reduced drift rates \citep[e.g.,][]{Johansen:2006p7466} or
pressure traps \citep{Pinilla:2012p16999} are needed to reconcile theory and observations. To
account for these open issues, we will also present models with reduced or without radial drift and
without fragmentation (see Sect.~\ref{sec:results:growth_only}).

\section{Results}\label{sec:results}
In this section we will discuss the results of two parameter studies: one of entirely MRI-active
disks with varied levels of turbulence, and another which mimics MRI-inactive regions (dead
zones) in which we vary the turbulence in both the MRI-active and dead regions
as well as the efficiency of radial drift. A summary of the models and parameters is given in
Table~\ref{tab:initial_conditions}.

\begin{table*}
\caption{Parameters and initial conditions of the simulations}
\label{tab:initial_conditions}
\centering
\begin{tabular}{l c c c D{.}{.}{-1} c c}
\hline\hline
Model &  $\alpha_\mathrm{A}$ & $\alpha_\mathrm{D}$ & $M_\mathrm{disk}$ [$M_\odot$] &  \multicolumn{1}
{c}{$\uf$ [m s$^{-1}$]} & \Rd [AU] & $\Edr$ \\    
\hline
\texttt{A2        } & $10^{-2}$ & $10^{-2}$ & 0.20 & 3.0  & -  & 1.0\\
\texttt{A3        } & $10^{-3}$ & $10^{-3}$ & 0.20 & 3.0  & -  & 1.0\\
\texttt{A4        } & $10^{-4}$ & $10^{-4}$ & 0.20 & 3.0  & -  & 1.0\\
\texttt{A5        } & $10^{-5}$ & $10^{-5}$ & 0.20 & 3.0  & -  & 1.0\\
\texttt{D23\_M05  } & $10^{-2}$ & $10^{-3}$ & 0.05 & 3.0  & 35 & 1.0\\
\texttt{D25\_M05  } & $10^{-2}$ & $10^{-5}$ & 0.05 & 3.0  & 35 & 1.0\\
\texttt{D35\_M05  } & $10^{-3}$ & $10^{-5}$ & 0.05 & 3.0  & 35 & 1.0\\
\texttt{LDE       } & $10^{-2}$ & $10^{-5}$ & 0.05 & 3.0  & 35 & 0.1\\
\texttt{COAG      } & $10^{-5}$ & $10^{-5}$ & 0.20 & $-$  & -  & 0.0\\
\hline
\end{tabular}
\tablefoot{Models \texttt{A2} through \texttt{A5} are the completely active disks, models starting
with \texttt{D} are the disks with dead zones, \texttt{LDE} stands for low drift efficiency,
and \texttt{COAG} is a model with neither radial drift nor fragmentation.
Other relevant initial conditions are the stellar mass $M_\star= 1\,M_\odot$, stellar
temperature $T_\star = 4300$~K, and $R_\star = 2.5$~$R_\odot$.}
\end{table*}

\begin{figure*}[ptb]
  \centering
  \resizebox{\hsize}{!}{\includegraphics{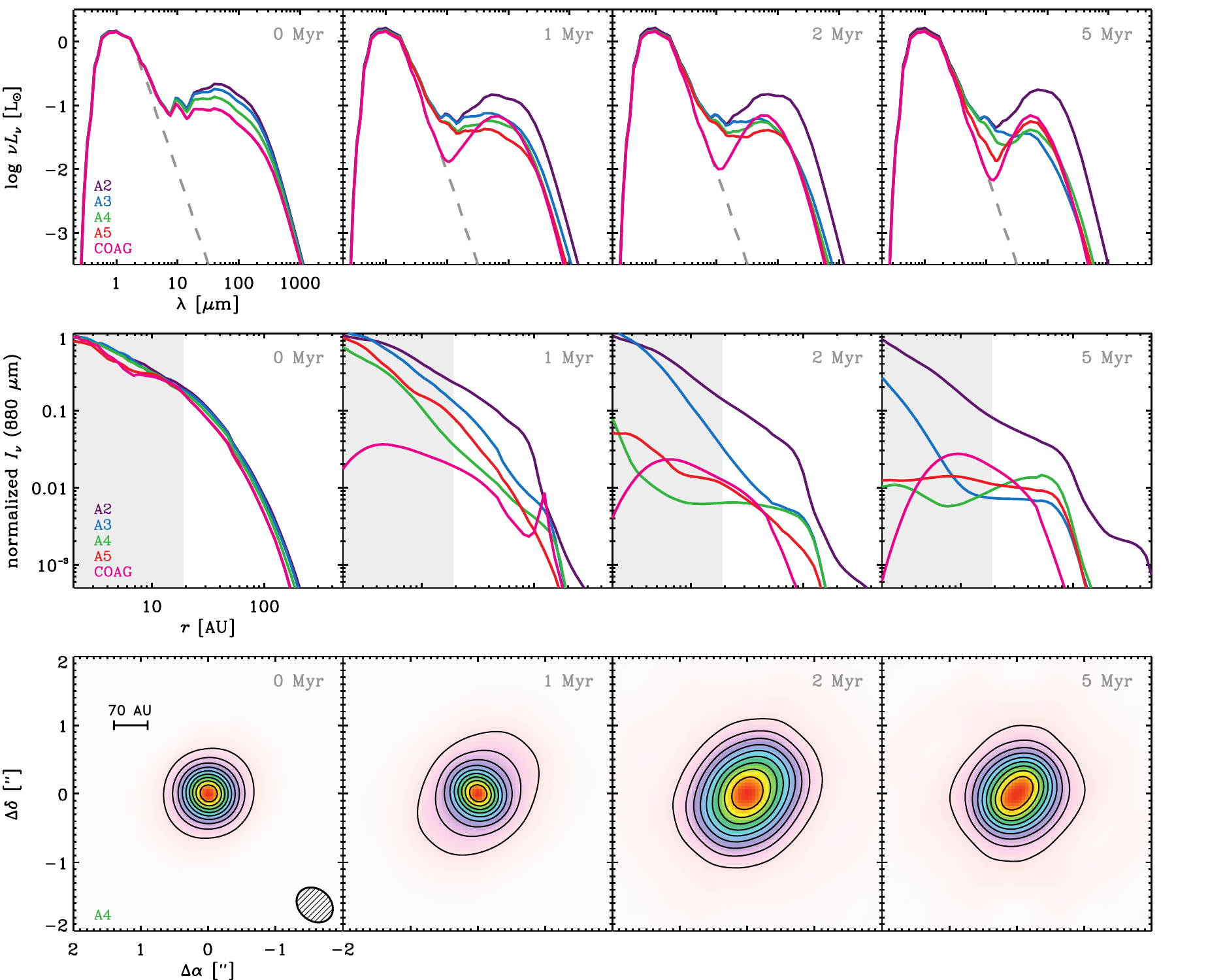}}
  \caption{Spectral energy distributions (top row), normalized radial surface brightness profiles at
  880~$\mu$m (middle row), and synthesized 880~$\mu$m images (bottom row) corresponding to the 
  Submillimeter Array 
  setup of \citet{Andrews:2011p16142} for the active disk models. The line colors correspond to
  models \texttt{A2} to \texttt{A5} and model \texttt{COAG}, as labeled in the left
  panels. The dashed grey line in the SEDs represents the stellar photosphere. The columns from left
  to right correspond to 0, 1, 2, and 5 Myrs of evolution. The shaded areas in the central row
  depict the resolution limits of current (pre-ALMA) interferometers. Only the results for model
  \texttt{A4} are shown in the synthesized images, where the contours correspond to 10\% levels of
  the peak brightness and the bar and dashed oval represent the size scale and the beam size,
  respectively.
  }
  \label{fig:SED_active}
\end{figure*}

\begin{figure*}[ptb]
  \centering
  \resizebox{\hsize}{!}{\includegraphics{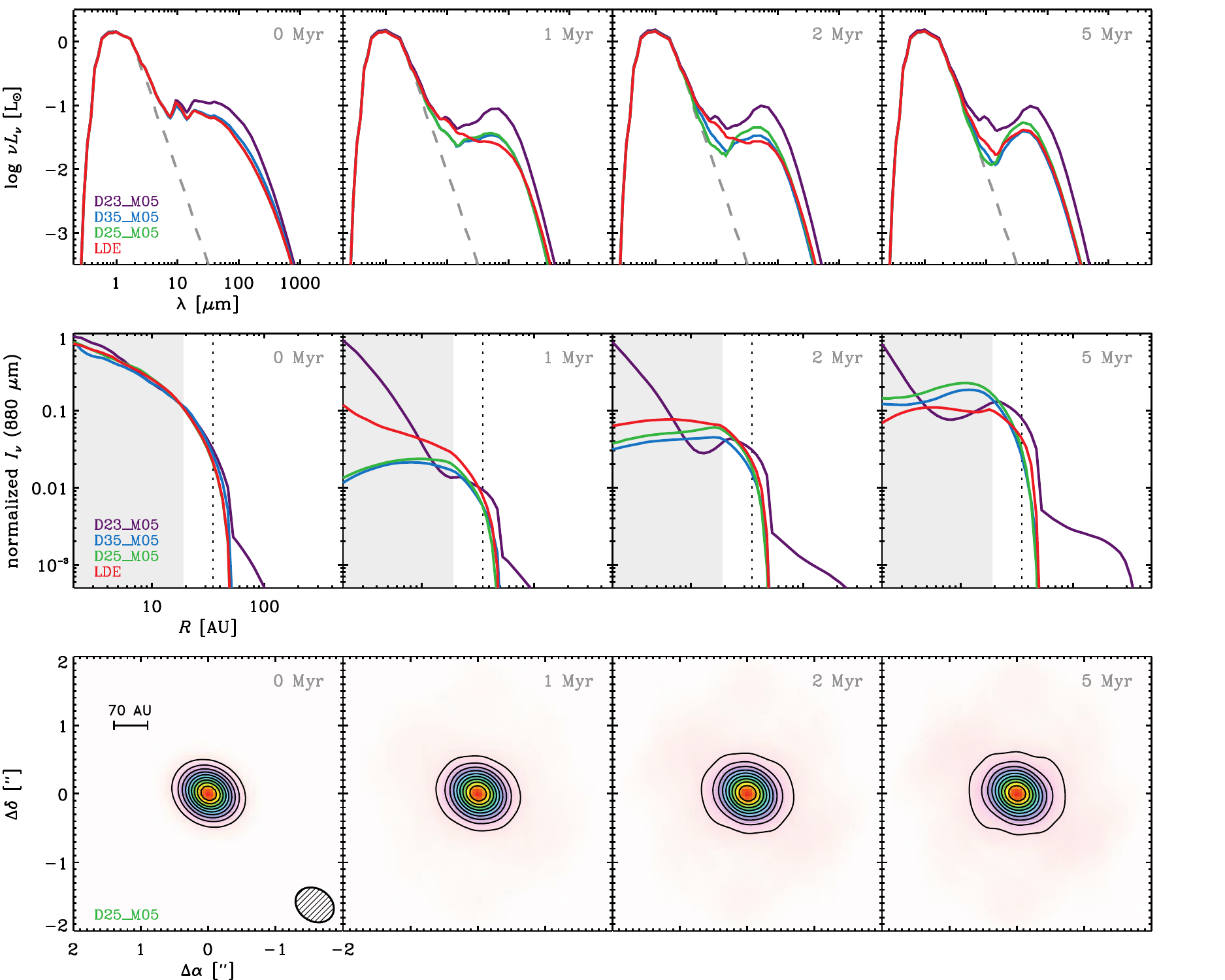}}
  \caption{Same as Fig.~\ref{fig:SED_active} but for the models which include
  dead-zones, corresponding to the simulations labeled in the left panels. The dashed vertical line in the
  central row marks the dead zone radius \Rd. The synthesized images in the bottom row
  correspond to model \texttt{D25\_M05}.}
  \label{fig:SED_dead}
\end{figure*}

\subsection{Growth and transport in smooth disk profiles}\label{sec:results:growth_only}
\subsubsection{Evolution of the dust distribution}
In this first parameter study, we investigate how the possible transition from a fragmentation
dominated ($a_\mathrm{f}<a_\mathrm{d}$) to a drift dominated ($a_\mathrm{d}<a_\mathrm{f}$) size
distribution is reflected in the spectral energy distribution (SED) of a typical disk profile. To
this end we set up different simulations with the same initial conditions and vary the constant
turbulence parameter \alphat from $10^{-2}$ to $10^{-5}$ in simulations \texttt{A2} to \texttt{A5},
respectively. According to Eq.~\ref{eq:a_frag}, this variation in \alphat shifts the fragmentation
barrier with respect to the drift barrier (Eq.~\ref{eq:a_drift}) by three orders of magnitude in
size. Thus the maximum size of the dust grains in the simulation with the highest \alphat value of
$10^{-2}$ (i.e., simulation \texttt{A2}) is set by fragmentation, while the drift limit becomes more
important as we go to lower turbulence values. In this parameter study, we vary only the turbulence
parameter, however it is important to note that the same or similar effects can be achieved by
varying the initial dust-to-gas mass ratio \Sigdust/\Siggas (for $\Sigdust\ll\Siggas$), or the
fragmentation velocity \uf, since these parameters determine the initial ratio of the size limits
(Eqns.~\ref{eq:a_frag} and \ref{eq:a_drift}). Changing these parameters, a drift-limited size
distribution can be achieved, even for a disk with relatively strong turbulence.

Figure~\ref{fig:contours_active} shows $\sigma(r,a)$, the dust surface density distributions as
function of grain size and radius for the active disk simulations, \texttt{A2} through \texttt{A5},
after 5~Myrs of evolution. The quantity $\sigma(r,a)$ is defined as
\begin{equation}
\sigma(r,a) = \int_{-\infty}^\infty n(r,z,a) \, m \, a \, \mathrm{d}z,
\label{eq:sigma_dust}
\end{equation}
where $m=4\pi/3\,\rhos\,a^3$ is the particle mass, and $n(r,z,a)$ is the dust number density
distribution as a function of radius $r$, height above the mid-plane $z$, and grain size $a$.
Integration of $\sigma(r,a)$ over $\ln(a)$ gives the dust surface density $\Sigdust(r)$. The solid
red line in Fig.~\ref{fig:contours_active} denotes the fragmentation limit (cf.
Eq.~\ref{eq:a_frag}), the dashed red line the drift size limit (cf.
Eq.~\ref{eq:a_drift}).
 
Going from simulation \texttt{A2} to \texttt{A5} (see Fig.~\ref{fig:contours_active}), it is 
evident that the small grain population becomes increasingly depleted. The reason for this behaviour is
the following: radial drift causes the dust surface density to be decreased on shorter time scales
than the gas surface density. The dust-to-gas ratio is therefore decreasing with time. The ratio of
the two growth barriers, which in this study is initially set by the choice of \alphat, therefore
shifts with time from a fragmentation limited distribution towards a drift limited one.  As
described in \citetalias{Birnstiel:2012p17135}, for a drift limited size distribution, dust collision
velocities induced by radial drift decrease with the dust-to-gas ratio. Therefore, at later times,
the low collision velocities prevent grain fragmentation. Consequently, small dust particles are not
reproduced by fragmentation anymore, but are rather swept up by the inward-drifting population of
larger grains.

The time scale of this sweep-up can be estimated in the context of the drift-dominated
two-population model of \citetalias{Birnstiel:2012p17135} (see Appendix~\ref{sec:appendix}) as
\begin{eqnarray}
\tau_\mathrm{D} &\simeq&  600~\mathrm{yr} \, \frac{r}{\mathrm{AU}} \cdot \frac{2.75}{\gamma}\,
\left(\frac{\epsilon}{10^{-2}}\right)^{-1}\,\left(\frac{T}{100~\mathrm{K}}\right)^{-\frac{1}{2}}\, \\
\tau_\mathrm{T} &\simeq& 1100~\mathrm{yr} \, \frac{r}{\mathrm{AU}} \cdot \frac{2.75}{\gamma}\,
 \left(\frac{\epsilon}{10^{-2}} \cdot \frac{T}{100~\mathrm{K}} \cdot \frac{\alphat}{10^{-3}}\right)^{-\frac{1}{2}}
\label{eq:tau_sweepup}
\end{eqnarray}
where $\epsilon$ is the dust-to-gas mass ratio, $\tau_\mathrm{D}$ and $\tau_\mathrm{T}$ are the
sweep-up time scales for drift and turbulence induced collision velocities. The dependence on
temperature $T$ and $\gamma$ is not relevant, considering that these values do not change
drastically with time or radius and the dependence on $T$ is weak. The sweep-up time scale is
therefore set by the quantities $r$, $\epsilon$, and \alphat. The late stages of disk evolution
after 5~Myrs in Fig.~\ref{fig:contours_active} are typically paired with dust-to-gas
mass ratios of $10^{-4}$ to $10^{-5}$, quite irrespective of $\alphat$.\footnote{As long
as the dust distribution becomes drift dominated, this leads to a self-regulating process: radial
drift removes dust mass, thus lowering the dust-to-gas ratio.
As the dust-to-gas ratio drops, the growth time scale and therefore also the drift time scale
increases. Hence, the evolutionary time scale of the dust-to-gas ratio depends on the dust-to-gas
ratio itself, and to reach a dust evolution time scale of a few Myrs, the dust-to-gas ratio needs to
be in the range of $10^{-5}$ to some $10^{-4}$.} The radial dependence and time
evolution of the dust to gas ratio is very similar to the results shown in Fig. 8 of
\citetalias{Birnstiel:2012p17135}. Therefore, the sweep-up time scales approach $\sim$Myrs values 
beyond about 10~AU. This explains why dust at these radii is not swept up
quickly. The linear dependence with $r$ means that the clearing of dust in this picture proceeds
from inside-out, as typically associated with other disc-clearing mechanisms
\citep[e.g.,][]{Ercolano:2011p14063}. In the innermost regions, the drift size limit approaches the
fragmentation limit. This causes some fragmentation of the largest particles, which is the source of
the small dust population inside of ~$\sim$0.5~AU for simulation \texttt{A5} and inside of
$\sim$1~AU for simulation \texttt{A4}.

\subsubsection{Evolution of the SED}
The resulting SEDs and 880~$\mu$m intensity profiles for all the MRI-active disk simulations are
shown in Fig.~\ref{fig:SED_active}, where the panels from left to right show the time evolution of
the different models. It can be seen that the mid-IR emission of \texttt{A2} and \texttt{A3} are
comparable up to about 20~$\mu$m, because even at 5~Myrs the fragmentation barrier in both cases is
still at similar or even smaller grains sizes than the drift limit. Going to the lower \alphat
values of \texttt{A4} and \texttt{A5} causes an ever stronger reduction of the mid-IR emission, due
to the fact that the drift limit becomes significantly smaller than the fragmentation limit and the
small dust grains in the hot inner regions are swept up by the drifting grains.

The outer regions show a similar effect: the emission beyond about 30~$\mu$m stays high in the case
of \texttt{A2} due to the fact that the strong level of turbulence prevents the formation of larger
grains. The dust grains in this case are so small that they are well coupled to the gas. Hence, the
dust is not strongly drifting, which means that more mass is retained in the outer regions and the
emission therefore stays at a high level.  Reducing the turbulence strength in the fragmentation
limited case leads to larger particles which start to attain significant inward drift velocities.  
The far-IR flux is therefore decreased, because less dust mass is retained in the outer regions
of the disk. This explains the strong difference between the SEDs of model \texttt{A2} compared to
the other models.

\subsubsection{Evolution of the (sub-)mm brightness}
The shape of the final radial intensity profiles in the 10 - 100~AU range can be understood by
approximating $I_\nu \,\propto\, T\,\Sigdust\,\kappa_\nu$. For simulations \texttt{A3}, \texttt{A4},
and \texttt{A5}, the largest grain size between about 7 and 50~AU is given by the drift-limit
Eq.~\ref{eq:a_drift} and at the same time is larger than $880~\mu$m$/(2\pi)$. This means that the
opacity \kap is roughly $\kap \, \propto\, a_\mathrm{dr}^{-1} \, \propto \, T \,\gamma\,r/\Sigdust$.
The intensity then becomes $I_\nu \propto T^2 \, \gamma \, r$, which for a temperature profile
$T\sim r^{-1/2}$ leads to a flat intensity profile. The entire disk of model \texttt{A2} is
fragmentation-dominated. This gives rise to a steeper intensity profile because the presence of
large amounts of small dust means the opacity does not depend as strongly on the largest grain
size. For a typical size distribution in the fragmentation case, we get $I_\nu \propto
\sqrt{\Siggas/T}$ which explains the slope of model \texttt{A2}.

\subsubsection{Conclusions for the smooth disk models}
Even though the SED signatures of the low turbulence simulations do resemble the ones of typical
transition disks, there is an important drawback of this mechanism when it comes to disks with
resolved holes in the \mbox{(sub-)}millimeter wavelength range: as pioneered by
\citet{Brown:2009p8895} and confirmed by other studies such as \citet{Andrews:2011p16142}, many of
the transition disks (but also some of the disks without typical transition disk-like SEDs) were
found to have inner dust cavities with sizes up to 70~AU. While it is easy to ``hide'' such cavities
in the SED by relatively small amounts of small dust, the resolved optically thin
\mbox{(sub-)}millimeter observations unambiguously reveal these features,
as shown e.g., by \citet{Isella:2010p17527} and \citet{Andrews:2011p16142}.

The mechanism presented here is able to remove the small dust population; however, it does so by
locking it up in a non-fragmenting population of larger dust particles. This population of grains
contains still enough particles of $\sim$centimeter sizes such that they clearly show up in the
\mbox{(sub-)}millimeter images. So even in the best case (model \texttt{A4}), no cavity is seen in
the simulated images (Fig.~\ref{fig:SED_active}, bottom row). The size of the largest particles
scale with the gas surface density; however, we chose a very high disk mass (20\% of the mass of the
central star) in order to reach the largest possible sizes for the given disk model. Even in this
case, the images shown in Fig.~\ref{fig:SED_active} do not resemble the observational data of e.g.,
\citet{Brown:2009p8895}. In the following section, we will investigate how far this mechanism can be
pushed by including effects which might be caused by a layered structure of the disk.

\subsection{Beyond standard dust modeling}\label{sec:results:dead_zone}
\subsubsection{Including effects of a layered disk structure}
So far, we have shown that the mechanism presented in this paper is able to strongly deplete the
inner regions of disks in small dust grains and that a mid-IR dip in the SED, as found by
observations
\citep[e.g.][]{Skrutskie:1990p16132,Calvet:2002p10424,Espaillat:2007p17013,Espaillat:2010p17008} can
be produced. In order to resemble the resolved (sub-)mm observations such as
\citet{Brown:2009p8895}, \citet{Andrews:2011p16142}, or \citet{Lyo:2011p16794}, two further features
are required.  First, the dust cavities also need to be observed in the (sub-)mm, i.e. for larger grains.
The absence of a cavity in the simulated images of Fig.~\ref{fig:SED_active} indicates that the
grain sizes found in our simulations are not large enough. Larger surface densities or larger
fragmentation thresholds can be considered. The former option, however, is restricted by the total
mass of the disk. Second, in our simulations, the dust surface density and the largest grain sizes
are smooth functions  of the radius, giving rise to a flat intensity profile (cf. the
estimate in Sect.~\ref{sec:results:growth_only}). This is in contrast with the observed dust
cavities. These facts motivate a model which allows for higher gas surface densities and steep
changes in the maximum grain size at the cavity edge.

One candidate for this behavior could be disks with dead zones, regions of low ionization which are
not MRI-active. The inactive regions transport angular momentum less efficiently
\citep{Armitage:2011p16081}, which is why they can accumulate large amounts of mass, while the
outer, active regions still resemble normal disks. This way, the inner, MRI-dead regions can have
higher surface densities, while the transition from MRI-active to MRI-dead gives rise to strong
changes in the turbulence parameter \alphat \citep[e.g.][]{Dzyurkevich:2010p11360}.

In this section, we therefore take these two effects into account without any detailed modeling of
the dead zone. This means that neither the vertical structure nor the viscous evolution of the dead
zone are considered; only the reduced \alphat and the increased gas surface density \Siggas are
treated. This is a valid approach because we are considering this as a best-case model. Other
effects such as fragmentation and enhanced radial mixing in the active surface layers are working
against the presented mechanism by replenishing small dust grains.

The situation where \alphat is higher in the outer parts of the disk and lower inside the dead zone
can cause the dust distribution to be fragmentation limited outside and drift limited inside the
dead zone. Models \texttt{D23\_M05} and \texttt{D25\_M05} simulate such a situation for $\Rd =
35$~AU, but for different contrasts between the active and the dead region. The size of the dead zone is
not physically motivated, but rather chosen from the typical size of the observed transition disk
cavities. However, the theoretical models of \citet{Bai:2011p16748} can account for dead zones of
sizes up to 20~AU.

A comparison of the SEDs for model \texttt{D23\_M05} (Fig.~\ref{fig:SED_dead}, purple SED) and \texttt{A2}
(Fig.~\ref{fig:SED_active}, purple SED) highlights how the contrast of one order of magnitude in \alphat
changes the long wavelength emission in the SED.  For our choice of parameters, the outer
regions start to be effected by drift, but fragmentation stays active throughout the disk.
Only at 5~Myrs, between around 10~AU and the outer edge of the dead zone, the drift barrier drops
slightly below the fragmentation barrier, causing a minor reduction of small grains.

We also compared models with different disk masses (not shown), but no significant
differences were found for changes within a factor of a few. Going to very low disk
masses of 0.005~$M_\odot$ further reduces the maximum grain size, which is producing relatively more
mm- and IR emission in the inner regions. Severe changes are found if the contrast in
\alphat between the active and the dead regions is increased, because this changes the relative
importance of the fragmentation and the drift barrier, as already discussed in the previous section.
The simulated images in Fig.~\ref{fig:SED_dead} (bottom row), however, do not show any signs of a
cavity. The reason for this is that the peak of the size distribution reaches sizes of around a
centimeter and larger only within a few AU (see third plot from top of
Fig.~\ref{fig:contours_dead}). Therefore, the surface density of dust smaller than $\sim$~1~cm only
decreases smoothly inwards of 3-8~AU, depending on the age of the disk.

In order to confirm this trend, we carried out simulations in which the size limits are pushed to even 
larger values. One reason for larger grains can be the efficiency of radial drift:
simulations by \citet{Johansen:2006p7466} and \citet{Bai:2010p15702} showed that the radial drift
velocity in the case of a turbulent environment can be reduced compared to the laminar, single-size
results of \citet{Weidenschilling:1977p865} and \citet{Nakagawa:1986p2048}. In order to artificially
change the efficiency of radial drift, we have introduced the efficiency factor \Edr in
Eq.~\ref{eq:gamma}. From Eq.~\ref{eq:a_drift}, it follows that $a_\mathrm{drift}$ then becomes inversely
proportional to the drift efficiency. This means that if radial drift is only $\Edr=0.1$ times as
fast as the classical value, $a_\mathrm{frag}$ became $1/\Edr=10$ times larger\footnote{It should be
noted that in the fragmentation dominated limit ($a_\mathrm{frag}<a_\mathrm{drift}$), the drift
velocity of the largest particles is decreased if \Edr is decreased while the particle size does not
change. In the drift limit ($a_\mathrm{drift}<a_\mathrm{frag}$), the situation is reversed: the
largest particles will grow to larger sizes until their drift time scale again equals the growth
time scale. Thus the particle size is increased while the drift velocity of the largest particles is
unchanged as long as their Stokes number is below unity.}.

Our simulation results with reduced drift efficiency (cf. model \texttt{LDE} in
Table~\ref{tab:initial_conditions}) show that particles do reach larger sizes in the dead zone (see
third panel from top in Fig.~\ref{fig:contours_dead}). Still, the grain size is a smooth
function of radius inside the dead zone. We found that going from 20 to 1~AU, the total surface density of grains
smaller than one centimeter decreases only smoothly by two orders of magnitude. This means that the
cavities in the mm-images cannot be explained with this mechanism. In addition to that, the reduced
drift speed also increases the time scale on which small dust is swept up (cf.
Eq.~\ref{eq:tau_sweepup}). This results in much larger amounts of small dust being left over, as can
be seen by comparing the second and third panel in Fig.~\ref{fig:contours_dead}.  It
also explains the less pronounced infrared dip in the SEDs (cf. Figs~\ref{fig:SED_active} and
\ref{fig:SED_dead}).

The intensity shapes inside the dead zones can be understood as in the previous subsection: for all
models but \texttt{D23\_M05}, we recover a rather flat intensity profile due to the presence of
large grains and a drift limited distribution. In model \texttt{D23\_M05}, the effective
fragmentation causes a steep increase in the inner regions, as in model \texttt{A3}. This increase
is mostly caused by changes in the opacity due to the presence of large amounts of small dust.

\subsubsection{Growth through all barriers}
It is a valid question to ask what disks would look like if our current understanding of
 the growth barriers and radial drift is entirely wrong. For this reason we simulated a case where
 grain growth is not inhibited by either fragmentation or radial drift. That is, we
 consider perfect sticking (without any fragmentation) and in addition, dust is only allowed to be
 diffused or dragged along by the gas, but not to drift (i.e., $\Edr = 0$). Apart from those
 changes, all other parameters of this model are identical to \texttt{A55} (see model \texttt{COAG} in
 Table~\ref{tab:initial_conditions}). The resulting grain size distribution is plotted in the bottom
 panel of Fig.~\ref{fig:contours_dead} and the resulting SEDs and brightness profiles are shown in
 Fig.~\ref{fig:SED_active}.

In Fig.~\ref{fig:SED_active}, it can be seen that for perfect sticking, the IR dip in the
SED forms very quickly and is very pronounced, as also found by \citet{Dullemond:2005p378}. While
the inner parts are cleared of small dust, the growing dust particles are not able to sweep up all
smaller dust further out in the disk, a population of small dust remains, as seen in the bottom
panel of Fig.~\ref{fig:contours_dead}. We calculated the time scale for the small dust sweep-up in
such a scenario (see Appendix~\ref{sec:appendix}), which is proportional to the planetesimal size and to the
quantity $r/(\Sigdust \csound)$ and for our setup exceeds 1~Myr beyond 10~AU. This approximation
does not include a reproduction of small dust, as would be expected for a distribution of
planetesimals, which would further increase the time scale. Therefore, even a scenario with perfect
sticking and no radial drift is not able to reproduce the observed large cavities in transition
disks.

\section{Discussion and conclusions}\label{sec:summary}
In this paper, we have investigated the ability of models of dust evolution to explain transition
disks. We calculated simulated SEDs and 880~$\mu$m images from the output produced by a dust
evolution code for a variety of different initial conditions. For clarity, we only considered
effects directly induced by growth, fragmentation and transport of dust, \emph{without} pressure
traps or planet induced gaps. The models were chosen to represent best-case scenarios for
this mechanism to work, i.e. large gas disk masses and no viscous evolution were assumed. The only
extensions to this model we have considered here are the ones which influence the grain size limits,
i.e., the strength of turbulence and the gas surface density. We found that effects of grain growth
can indeed produce dips in the SED such as found for many transition disks \citep[see
also][]{Dullemond:2005p378}, however they fail to reproduce the cavities in the
\mbox{(sub-)millimeter} images such as found by
\citet{Brown:2009p8895,Andrews:2011p16142,Lyo:2011p16794}.
Even in the extreme case where growth proceeds without any barriers and radial drift is
switched off, no \textit{large} inner cavities can be formed within 5~Myrs of evolution.

This leads to the conclusion that disks with large inner holes cannot be caused by grain growth
\textit{alone}. Due to the fact that current models of disk photoevaporation
\citep[e.g.][]{Owen:2010p14269} also fail to explain the inner cavities in the accreting objects, we
propose that a combination of dust evolution with other effects such as pressure bumps, spiral arms
or planet induced gaps could be the solution of the problem (see Pinilla et al., submitted to A\&A).
Important questions to be answered from the observational side are: apart from the size of the
cavity, is there is a distinct difference between disks with small and large cavities? What is the
gas content inside the cavities? And do different sizes of dust show different cavity
shapes in the same object \citep[see][]{Dong:2012p17966}? Theoretical models will need to investigate the
trapping mechanisms which effectively shepherds the dust outside the cavity, irrespective of the particle sizes.

Our findings can be summarized as follows:
\begin{itemize}
  \item A grain size distribution for which the drift-induced grain size limit is smaller than the
  fragmentation induced size limit becomes inefficient in replenishing small dust
  due to a lack of fragmenting collisions.
  \item The time scale on which the remaining small dust is swept up by the largest, inward drifting
  grains is proportional to the distance to the central star and depends also on the dust-to-gas
  mass ratio and the turbulence strength. The dispersal of small dust is therefore from the inside
  out.
  \item Small amounts of small dust can be retained for several million years in the outer parts of
  the disk ($\gtrsim 10$~AU) if the dust-to-gas mass ratio is lower than a few times $10^{-4}$.
  \item Even if the inner regions are cleared of small dust via grain growth, most of the dust mass
  is still detectable by millimeter observations. Furthermore, the grain size is a smooth function
  of the stellocentric radius which means that the ``observable'' dust surface density
  decreases only slightly and the brightness profile stays approximately flat. This
  mechanism thus fails to reproduce the strong drop in millimeter emission which is typically
  observed in disks with large cavities.
  \item The model does not resemble the observations of large cavities, even if perfect
  sticking and no radial drift is assumed. The observed, sharp division between the inner and the
  outer regions call for a severe change in the disk properties. Pressure bumps that are strong
  enough to decouple both large and small dust from the accretion flow are most likely necessary
  to explain the observations.
\end{itemize} 
\begin{acknowledgements} 
We like to thank Kees Dullemond, Chris Ormel, James Owen, and Satoshi Okuzumi for stimulating
discussions. We also thank the anonymous referee for his/her very constructive and helpful
criticism.
\end{acknowledgements}

\bibliographystyle{aa}
\bibliography{bibliography}

\appendix
\section{Estimation of the sweep-up time scales}\label{sec:appendix}
To estimate the time scale on which small dust is swept up by larger, drifting particles, we
consider the two-population model of \citetalias{Birnstiel:2012p17135} according to which the grain size
of the drifting particles is given by
\begin{equation}
a_1 = \frac{2\, f_\mathrm{d} \Sigdust \, V_\mathrm{K}^2}{\pi \, \rhos\,\gamma\,\csound^2},
\label{eq:app_a_drift}
\end{equation}
where $f_\mathrm{d}=0.55$ and $V_\mathrm{K}$ is the Keplerian velocity.
The drift velocity of the largest grains is then given by
\begin{equation}
u_1 = f_\mathrm{d} \,\epsilon\,V_\mathrm{K},
\end{equation}
where $\epsilon$ is the dust-to-gas mass ratio. Assuming that the large particles contain most of
the mass of the system ($\Sigma_1 \simeq \Sigdust$), and that the small particles are well mixed with
the gas ($\St_1\lesssim \alphat$), we can write the change of the surface number density of small
particles $N_0$ as \citep[see,][Appendix A.2.]{Birnstiel:2010p9709}
\begin{equation}
\dot N_0 = \frac{1}{\sqrt{2\,\pi}\,H_\mathrm{g}}\, u_{01} \, \sigma_{01} \, N_0 \, N_1.
\label{eq:app_n_dot}
\end{equation}
Approximating the relative velocity $u_{01}$ between the two species by the drift velocity of the
large grains and the collision cross section by $\sigma_{01}\simeq\pi a_1^2$, we can derive the time
scale for the sweep-up of small dust by drift-induced collisions
\begin{equation}
\tau_\mathrm{D} = \frac{N_0}{\dot N_0} = \frac{8}{3}\, \sqrt{\frac{2}{\pi}} \,\frac{r}{\epsilon \, \csound \, \gamma}. 
\label{eq:app_sweepup_d}
\end{equation}
The same calculation but using the turbulent relative velocities between the largest and smallest grains
\begin{equation}
u_{01} \simeq \frac{3}{2}\,\sqrt{\frac{\alphat\,a_1\,\rhos\,\pi}{\Siggas}}
\end{equation}
according to \citet{Ormel:2007p801} yields the sweep-up time of small particles where collisions are driven by turbulence,
\begin{equation}
\tau_\mathrm{T} \simeq \frac{r}{\csound}\, \sqrt{\frac{f_\mathrm{d}}{\alphat\,\epsilon\,\gamma}}.
\label{eq:app_sweepup_t}
\end{equation}
Comparing Eqns.~\ref{eq:app_sweepup_d} and \ref{eq:app_sweepup_t} shows that the sweep-up of the small dust is driven by turbulent velocities, as long as
\begin{equation}
\alphat \gtrsim \frac{2}{9}\,f_\mathrm{d}\,\gamma \cdot \epsilon, 
\end{equation} 
which is in our models approximately gives $\alphat \gtrsim \frac{\epsilon}{3}$.

A similar calculation can be done for the small dust in a distribution which is
dominated by planetesimals of size $a_1$. Using Eq.~\ref{eq:app_n_dot} with the cross section
$\sigma_{01}\simeq \pi a_1^2$ and the relative velocities from the head wind $\Delta u_{01}=
\csound^2\,\gamma/(2 V_\mathrm{K})$, we derive
\begin{equation}
\tau_\mathrm{P} = \frac{8\sqrt{2\pi}\rhos}{3}\, \frac{a_1\,r}{\Sigdust\,\csound\,\gamma}.
\label{eq:app_sweepup_planetesimals}
\end{equation}
For the values of our simulations (model \texttt{COAG}) and $a_1=10^{4}$~cm, we find that
$\tau_\mathrm{P} \simeq 1$~Myrs at 10~AU and increases steeply with radius. This agrees well with
the fact that the planetesimals cannot efficiently sweep up the small grains beyond a few AU, as
seen in the bottom panel of  Fig.~\ref{fig:contours_dead}.

\makeatletter
\if@referee
\processdelayedfloats
\pagestyle{plain}
\fi
\makeatother
\end{document}